\documentclass[namedreferences,hyperref,optionalrh,solaromanenum]{spr-sola}

\usepackage{graphicx}                    % For eps figures, newer & more powerfull
\usepackage{color}                       % For color text: \color command
\usepackage{natbib}
\usepackage{cleveref}
\usepackage{url}
\chardef\us=`\_

\begin{document}

\begin{frontmatter}

\title{Type I Solar Radio Bursts Modulated by Solar Flares}

%%%%%%%%%%%%%%%%%%%%%%%%%%%%%%%%%%%%%%%%%%%%%%%%%%%
%% Authors Names

\author[addressref=aff1,email={liyutong@dzu.edu.cn}]{\inits{Y.}\fnm{Yutong}~\snm{Li}}
\author[addressref=aff1,corref,email={licy@sdu.edu.cn}]{\inits{C.}\fnm{Chuanyang}~\snm{Li}}
\author[addressref=aff1]{\inits{Y.}\fnm{Yanke}~\snm{Tang}}
\author[addressref=aff1]{\inits{N.}\fnm{Ning}~\snm{Gai}}
\author[addressref=aff1]{\inits{Z.}\fnm{Zichuan}~\snm{Li}}
\author[addressref=aff1]{\inits{Z.}\fnm{Zhe}~\snm{Cui}}
\author[addressref={aff1,aff2}]{\inits{Y.}\fnm{Yang}~\snm{Gao}}
\author[addressref=aff1]{\inits{Y.}\fnm{Yifan}~\snm{Wang}}
\author[addressref=aff1]{\inits{X.}\fnm{Xiaodong}~\snm{Xu}}
\author[addressref=aff1]{\inits{X.}\fnm{Xiaodi}~\snm{Huo}}
\address[id=aff1]{Shandong Key Laboratory of Space Environment and Exploration Technology, College of Physics and Electronic Information, Dezhou University, Dezhou 253023,China}
\address[id=aff2]{Purple Mountain Observatory \& Key Lab. of Radio Astronomy, Chinese Academy of Sciences, Nanjing 210023, China}

%%%%%%%%%%%%%%%%%%%%%%%%%%%%%%%%%%%%%%%%%%%%%%%%%%%
%% Runningheads
%
\runningauthor{Y. Li et al.}
\runningtitle{Type I Solar Radio Bursts Modulated by Solar Flares}

%%%%%%%%%%%%%%%%%%%%%%%%%%%%%%%%%%%%%%%%%%%%%%%%%%%
%%% Abstract
\begin{abstract}
  Type I solar radio bursts (noise storms) are persistent meter-wave nonthermal emissions above active regions, with their occurrence and properties closely related to the local magnetic configuration and nonthermal electron acceleration. This study examines a type I noise storm on 24 December 2023 and its relation to flare activity. The noise-storm source was co-spatial with active region AR 3529 and showed frequency-dependent spatial dispersion. The associated M2.9 flare strongly modulated the emission, with the storm intensity decreasing at flare onset, recovering afterward, and shifting to higher frequencies. Based on multiwavelength observations, we suggest that pre-flare small-scale reconnection supplied nonthermal electrons to overlying closed magnetic structures and maintained the storm. During the flare, magnetic reconnection above the active region produced bidirectional plasma ejections and type III bursts with bidirectional frequency drifts; the gradually decreasing starting frequency of these bursts may indicate an upward-moving reconnection site. The resulting magnetic reconfiguration disrupted electron trapping and suppressed the storm, whereas post-flare magnetic recovery allowed the emission to resume. These results show that flares can modulate type I noise storms through magnetic restructuring and provide insight into the generation mechanism of noise storms.
\end{abstract}


\end{frontmatter}
%-------------------------------------------------

%%%%%%%%%%%%%%%%%%%%%%%%%%%%%%%%%%%%%%%%%%%%%%%%%%%
%% Sections
%
\section{Introduction}

Solar radio type I noise storms are among the most common meter-wave nonthermal radio emissions and can persist for hours to days in localized coronal magnetic structures above active regions. A typical dynamic spectrum consists of a slowly varying broadband continuum superposed by numerous short-duration, narrowband, pulse-like type I bursts. These bursts last 0.1--1 s, have relative bandwidths ($\Delta f/f_0$) of about 2\%--3\%, and often show nearly 100\% circular polarization in the ordinary (O) mode (\citealt{1946Natur.157...47H, elgaroy1977}). The emission frequency, intensity, and polarization depend sensitively on the local plasma density, magnetic-field strength, and field topology. Type I storms therefore serve as a useful diagnostic of coronal magnetic evolution, local disturbances, and the acceleration of low-energy nonthermal electrons in active regions (e.g., \citealt{1997ESASP.404..299C, 2000SoPh..193..227B, 2013ApJ...762...89R, 2015A&A...582A..52A, 2017SoPh..292...82L}).

Two mechanisms have been proposed for type I storm emission, and both depend on the magnetic and plasma conditions. The first and widely accepted mechanism is fundamental plasma emission (\citealt{1958SvA.....2..653G, 1970AuJPh..23..871M}). In this scenario, nonthermal electrons excite Langmuir waves in closed magnetic structures, after which scattering or nonlinear wave-wave interactions converts the Langmuir waves into type I radiation (e.g., \citealt{1980SoPh...67..357M, 1981A&A....94..100B, 1982A&A...105..221S}). This mechanism can account well for key observational characteristics of type I bursts, including frequency distribution, frequency-drift behavior, and high degree of circular polarization. The second mechanism, electron cyclotron maser emission (ECME; \citealt{1958AuJPh..11..424T, 1959PhRvL...2..504S, 1979ApJ...230..621W}), involves unstable electron distributions, such as loss-cone or horseshoe distributions, that can directly amplify electromagnetic waves under strongly magnetized or low-density conditions ($\omega_p/\Omega_e < 1$). ECME offers an alternative explanation for type I bursts with extremely high brightness temperatures and fine structures (e.g., \citealt{2012PhPl...19h2902W, tang2017ecm}).

Previous studies show that eruptive activity can strongly modulate type I storms, while the underlying processes remain unclear. Flares and coronal mass ejections (CMEs) often weaken or suppress type I emission (e.g., \citealt{1990SoPh..130...19A, 1993SoPh..145..151A, 1994scs..conf..271K, 2001SoPh..202..337C}).  \cite{1990SoPh..130...19A, 1993SoPh..145..151A} reported several flare-associated switch-off events. A statistical study by \cite{2001SoPh..202..337C} suggested that CMEs can open magnetic fields above active regions, allowing nonthermal electrons trapped in closed structures to escape into interplanetary space and thereby terminate type I emission. In contrast, eruptive activity can also enhance or initiate type I storms (e.g., \citealt{1994A&A...281..536R, 1996SoPh..167..333C, 1996SoPh..165..347M, 2012ApJ...744..167I}). \cite{2012ApJ...744..167I} proposed that side-lobe reconnection after CME eruption can continuously supply weak but persistent nonthermal electrons to sustain noise-storm emission, whereas subsequent loop expansion near the CME front can distort or disrupt the magnetic configuration and suppress the storm.

This study uses multiwavelength observations to examine flare modulation of a type I noise storm and the associated changes in the source environment. Section 2 describes the instruments and data sets, Section 3 presents the observational results, and Section 4 presents the summary and discussion.

\section{Instruments and Data}

ORFEES (Observations Radiospectrographiques pour FEDOME et l' Etude des Eruptions Solaires) provides solar radio dynamic spectrum from 144 to 1004 MHz with a temporal resolution of 0.1 s (\citealt{2021JSWSC..11...57H}). Data from the Radio Solar Telescope Network (RSTN) San Vito station (\citealt{1981BAAS...13Q.553G}), provides long-term total-flux monitoring at several fixed frequencies and dynamic spectra. The San Vito dynamic spectrum covers 25--180 MHz, with a frequency resolution of about 0.125 MHz and a temporal resolution of 3 s; the flux measurements have a cadence of 1 s.

We also use radio imaging data from the Nancay Radioheliograph (NRH) to locate the type I burst source and trace its evolution during flare activity. NRH observes in the 150--450 MHz range at ten frequencies (150, 173, 228, 270, 298, 327, 360, 408, 432, and 445 MHz), with an angular resolution of about 2--6 arcmin (depending on the observing frequency and mode) and a time resolution of 0.25 s (\citealt{1997LNP...483..192K}).

The extreme ultraviolet (EUV) images come from the Atmospheric Imaging Assembly (AIA: \citealt{2012SoPh..275...17L}) aboard the Solar Dynamics Observatory (SDO: \citealt{2012SoPh..275....3P}). AIA provides full-disk EUV images in multiple passbands, with a spatial sampling of $0.6''$ pixel$^{-1}$ and a cadence of 12 s. The photospheric magnetic field is characterized using line-of-sight magnetograms from the Helioseismic and Magnetic Imager (HMI: \citealt{2012SoPh..275..229S}) on SDO. HMI magnetograms have a spatial sampling of $0.5''$ pixel$^{-1}$ and a cadence of 45 s.

\section{Observational Results}

 The event occurred on 24 December 2023. This section first presents an event overview, followed by an analysis of the type I storm properties and the effects of associated flare activity.

\subsection{Event Overview}

\begin{figure} %%%%%%%%%%%%%%%%%% FIGURE 1
\centerline{\includegraphics[width=1\textwidth,clip=]{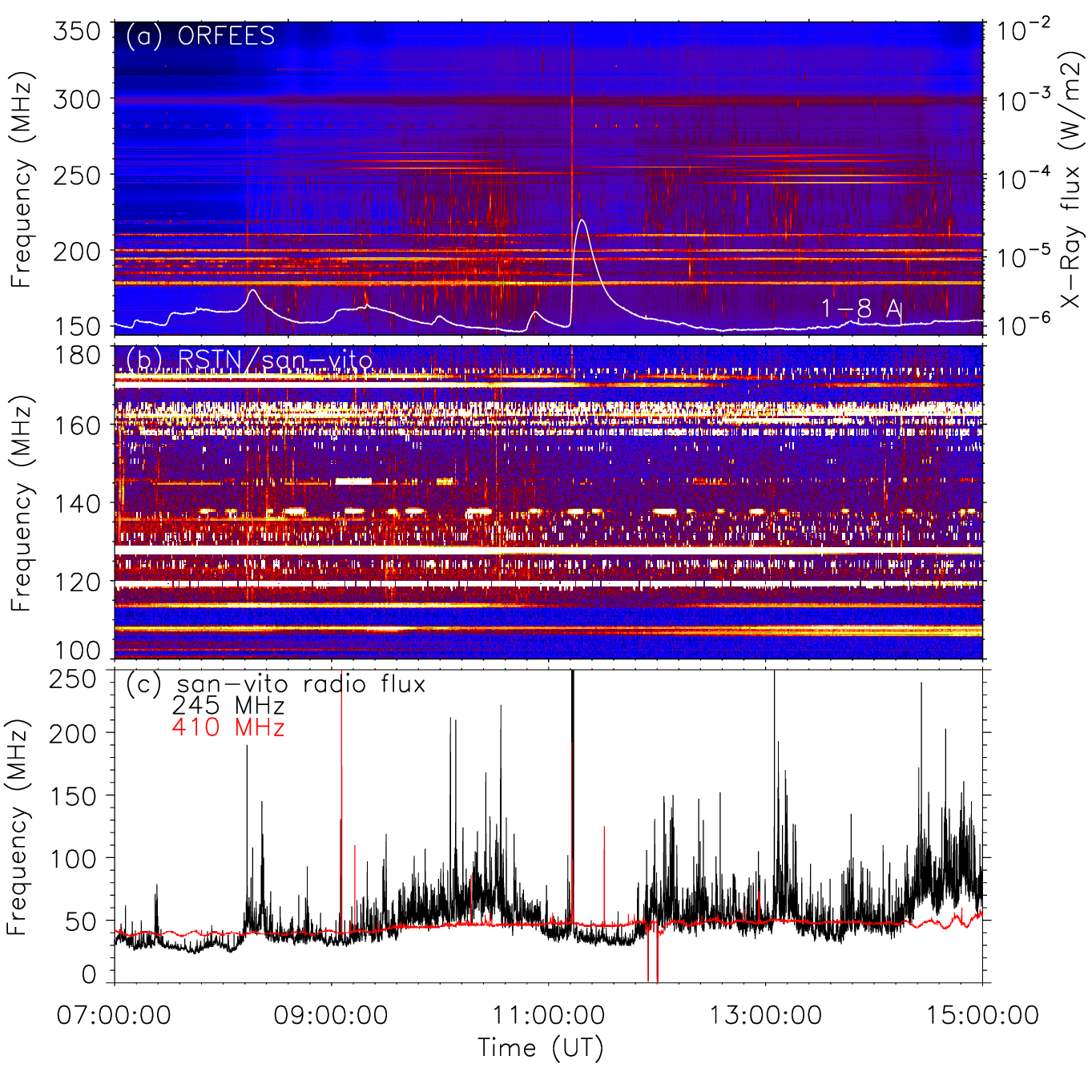}}
\small
        \caption{Solar radio dynamic spectra and radio fluxes from 07:00 to 15:00 UT on 24 December 2023. (a) ORFEES dynamic spectrum at 144--350 MHz, overlaid with the GOES-16 1--8 \AA\ soft X-ray light curve in white. (b) RSTN/San Vito dynamic spectrum at 100--180 MHz. (c) San Vito radio fluxes at 245 MHz (black) and 410 MHz (red).
                }
\label{Fig1}
\end{figure}

Figure \ref{Fig1} shows the radio dynamic spectra, radio fluxes, and GOES-16 soft X-ray flux from 07:00 to 15:00 UT. A long-duration type I noise storm appeared in the 100--350 MHz range, together with several broadband type III bursts. Three flare episodes occurred during this interval. The first began near 08:06 UT and peaked at 08:16 UT as a C3.8 flare, after which the type I emission gradually increased and became prominent. The second began at 10:43 UT and peaked at 10:52 UT as a C1.9 flare, followed by the strongest episode, an M2.9 flare, which began at 11:09 UT and peaked at 11:18 UT. No CME was reported in association with the M2.9 flare.

The M2.9 flare clearly modulated the type I storm. The type I emission began to decrease at approximately 10:25 UT, as indicated by the 245 MHz RSTN/San Vito radio flux. During the flare, the persistent type I emission was temporarily interrupted, and a type III burst occurred (Figures 1a and 1b). After the flare, the type I storm reappeared and remained detectable for several hours, with emission mainly above 150 MHz, whereas before the flare it extended further toward lower frequencies, down to about 120 MHz.

\begin{figure} %%%%%%%%%%%%%%%%%% FIGURE 2
\centerline{\includegraphics[width=1\textwidth,clip=]{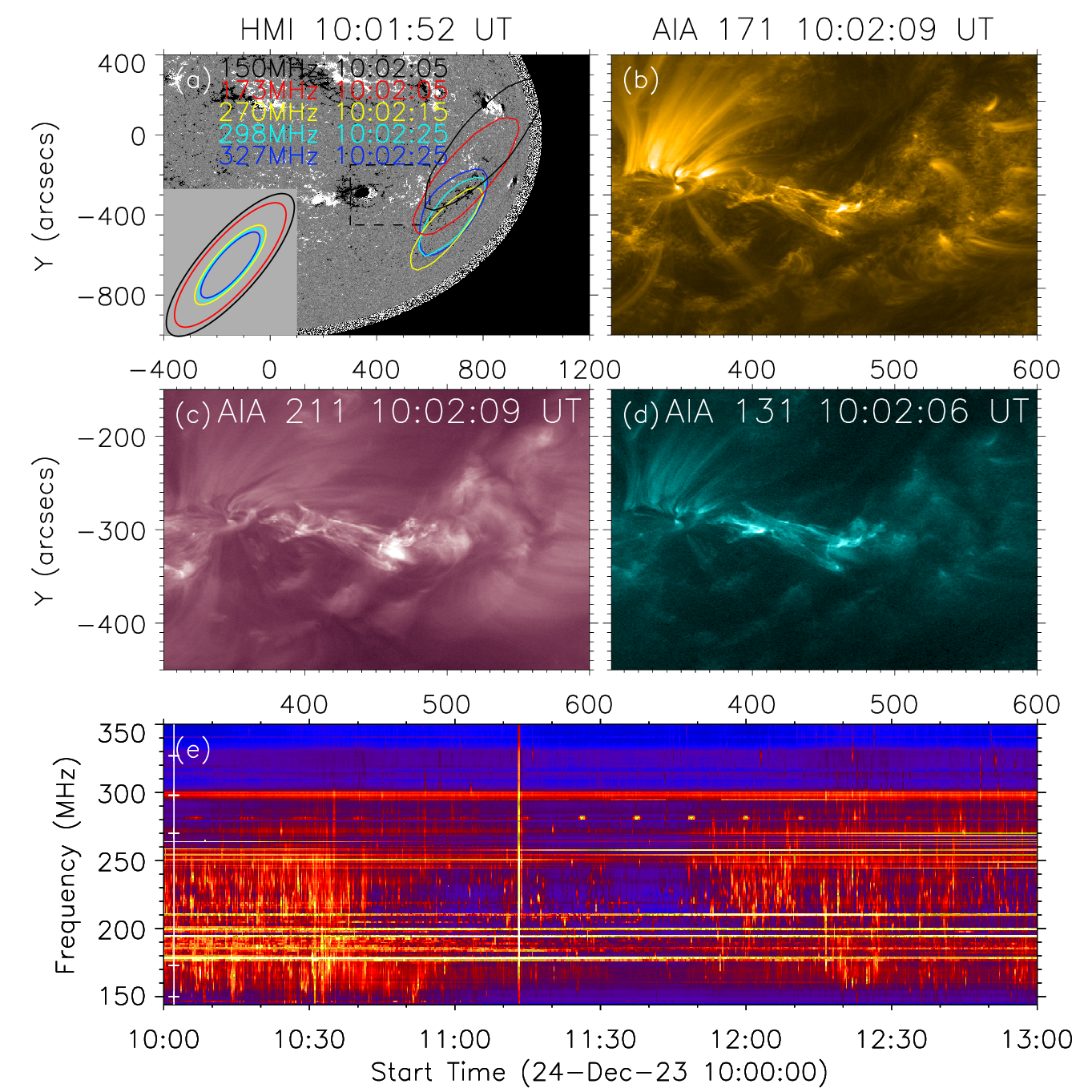}}
\small
        \caption{Magnetic-field and EUV images of the active region associated with the type I storm. (a) SDO/HMI line-of-sight magnetogram overlaid with NRH radio source positions at four frequencies, shown as 70\% peak-flux contours. The set of tilted ellipses in the lower-left corner shows the half-maximum levels of the NRH interferometric beam at different frequencies. (b)--(d) SDO/AIA 171, 211, and 131 \AA\ images of the region marked by the black dashed box in panel (a). (e) ORFEES radio dynamic spectrum. The white line and plus symbols mark the NRH imaging times and frequencies shown in panel (a). An animation accompanies this figure.
                }
\label{Fig2}
\end{figure}

Figure \ref{Fig2} presents magnetic field and EUV images of the active region associated with the type I storm. NRH multi-frequency imaging shows that the type I radio source remained spatially associated with AR 3529 throughout the event, although the accompanying animation reveals slight oscillations. The source show clear frequency-dependent spatial dispersion. During the flare, AIA images reveal prominent EUV jet activity, indicating magnetic reconnection and energy release. During the type III burst, the radio source shift northward, suggesting different electron transport paths and magnetic topologies for the type I and type III emissions.

\subsection{Type III Bursts with Bidirectional Frequency Drifts}

\begin{figure} %%%%%%%%%%%%%%%%%% FIGURE 3
\centerline{\includegraphics[width=1\textwidth,clip=]{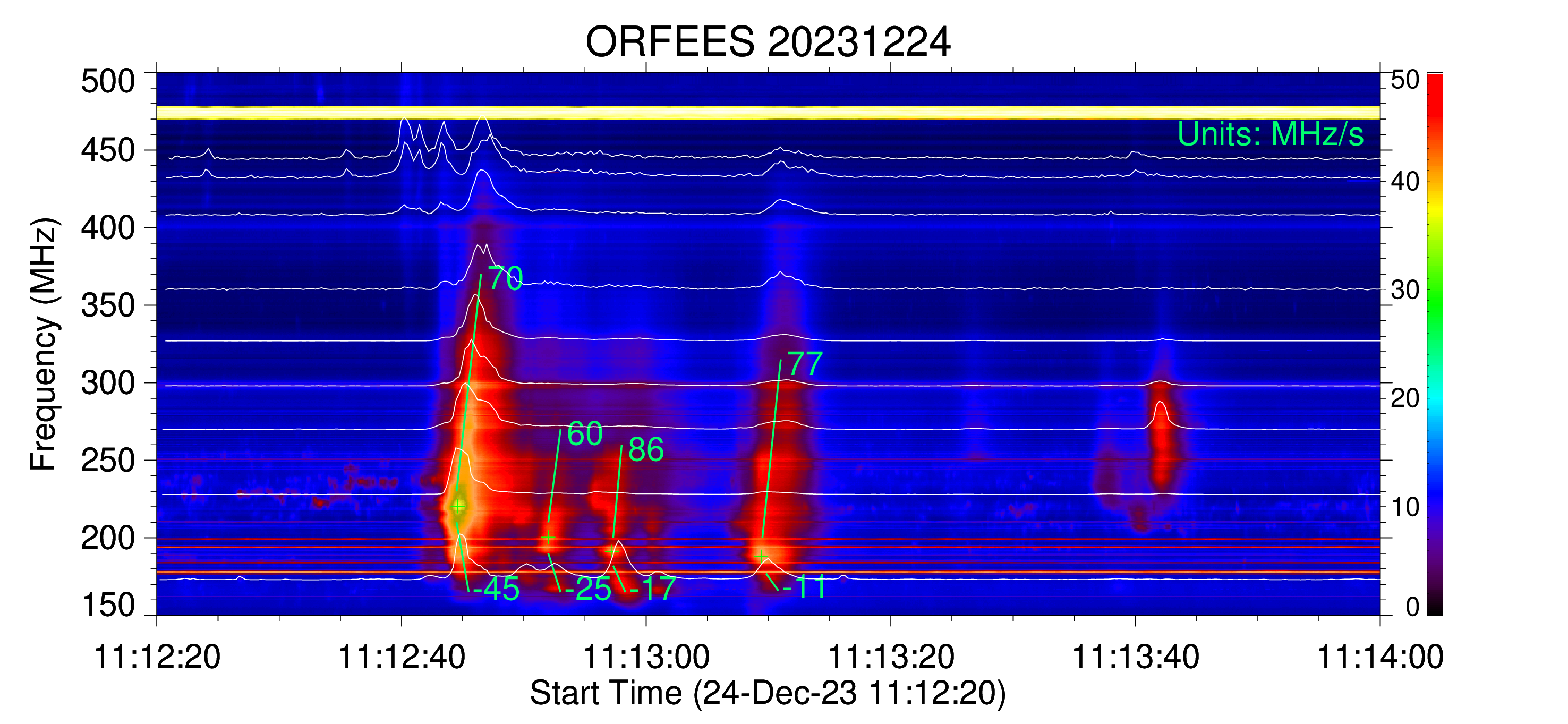}}
\small
        \caption{Zoomed-in view of the ORFEES dynamic spectrum from 11:12:20 to 11:14:00 UT at 150--500 MHz. White solid lines show the NRH radio fluxes at the corresponding frequencies. The numbers indicate the frequency drift rates of the bidirectional type III bursts, the straight line represents the linear fit, and the plus signs mark the starting (turning) points.
                }
\label{Fig3}
\end{figure}

Figure \ref{Fig3} shows a zoomed-in view of ORFEES dynamic spectrum from 11:12:20 to 11:14:00 UT. During this interval, a group of type III bursts with bidirectional frequency drifts appeared, with one branch drifting toward lower frequencies and the other toward higher frequencies. Such bursts provide radio signatures of magnetic reconnection, in which a localized reconnection site accelerates electron beams both upward and downward. Upward-propagating electrons travel along open or high coronal field lines, while downward-propagating electrons travel along lower field lines, producing opposite frequency drifts in the dynamic spectrum.

The normal-drifting branch had drift rates of about -11 -- -45 MHz s$^{-1}$, whereas the reverse-drifting branch had drift rates from 60 to 86 MHz s$^{-1}$. The starting frequency of the bidirectional burst decreased during the event (Figure \ref{Fig3} ), from 220.3 to 187.8 MHz between 11:12:44 and 11:13:09 UT, which may indicate a change in the source conditions. A few type I bursts appeared during this interval, but with much lower occurrence rates and intensities than before or after the flare. This suppression suggests that the flare and associated type III burst activity disrupted the type I emission process.

\subsection{Evolution of the Radio Source and Underlying Magnetic Fields}

\begin{figure}  %%%%%%%%%%%%%%%%%% FIGURE 4
\centerline{\includegraphics[width=1\textwidth,clip=]{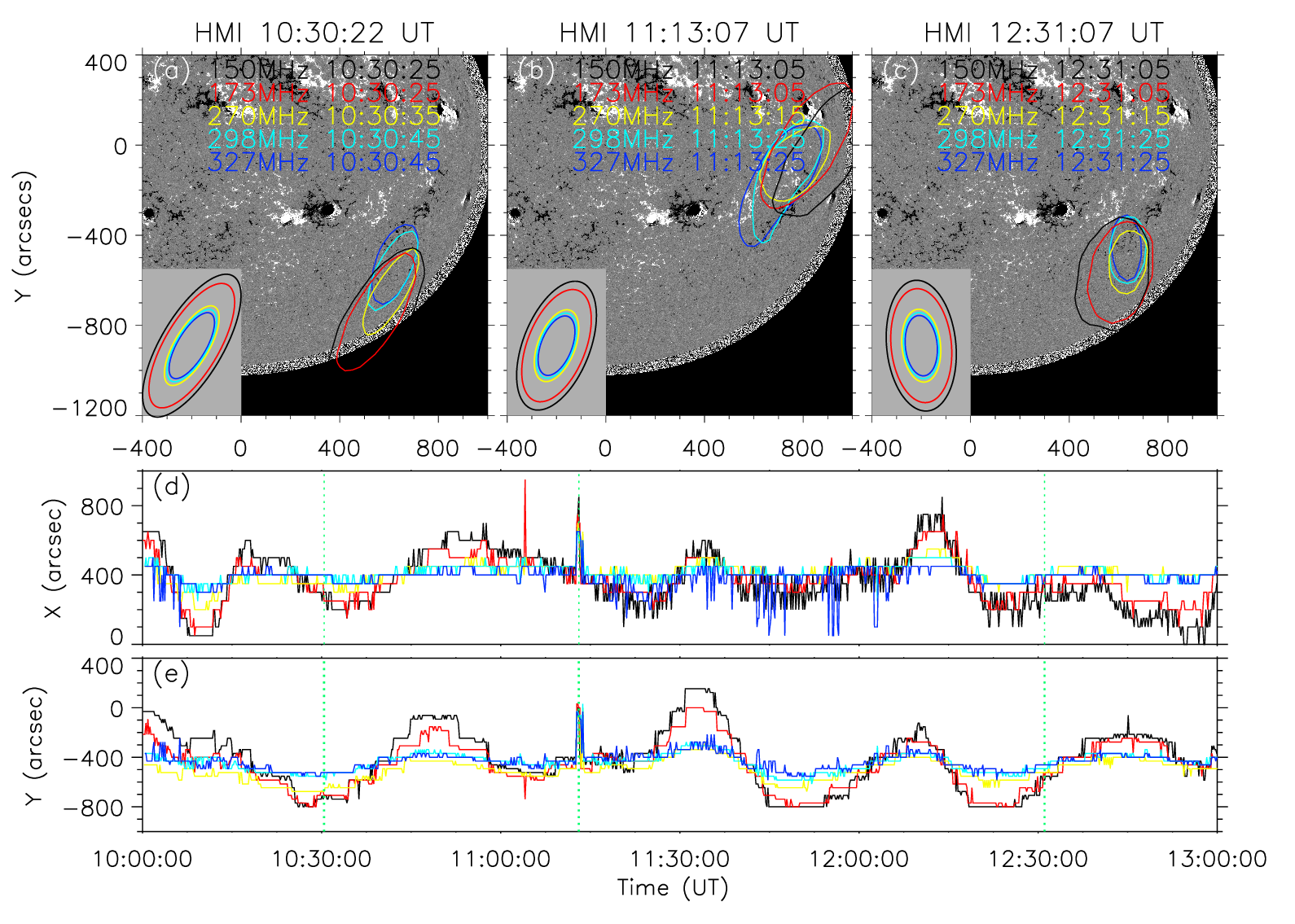}}
\small
        \caption{(a)--(c) NRH radio source positions before, during, and after the flare, overlaid on the corresponding HMI magnetograms. The set of tilted ellipses in the lower-left corner shows the half-maximum levels of the NRH interferometric beam at different frequencies. (d)--(e) Temporal evolution of the source positions (east--west and north--south) derived from the peak-flux locations.
         }
\label{Fig4}
\end{figure}

Figure \ref{Fig4} compares the source positions of the type I noise-storm before, during, and after the flare. NRH observations show no significant change in the projected source position before and after the flare, suggesting that the source region, or the conditions required for type I emission, recovered after the flare. Since this event occurred away from the limb and the radio source oscillated with time, the projected positions alone are insufficient to determine whether the height of the type I source region changed between the pre- and post-flare phases (Figures 4d and 4e).

\begin{figure}%%%%%%%%%%%%%%%%%% FIGURE 5
\centerline{\includegraphics[width=1\textwidth,clip=]{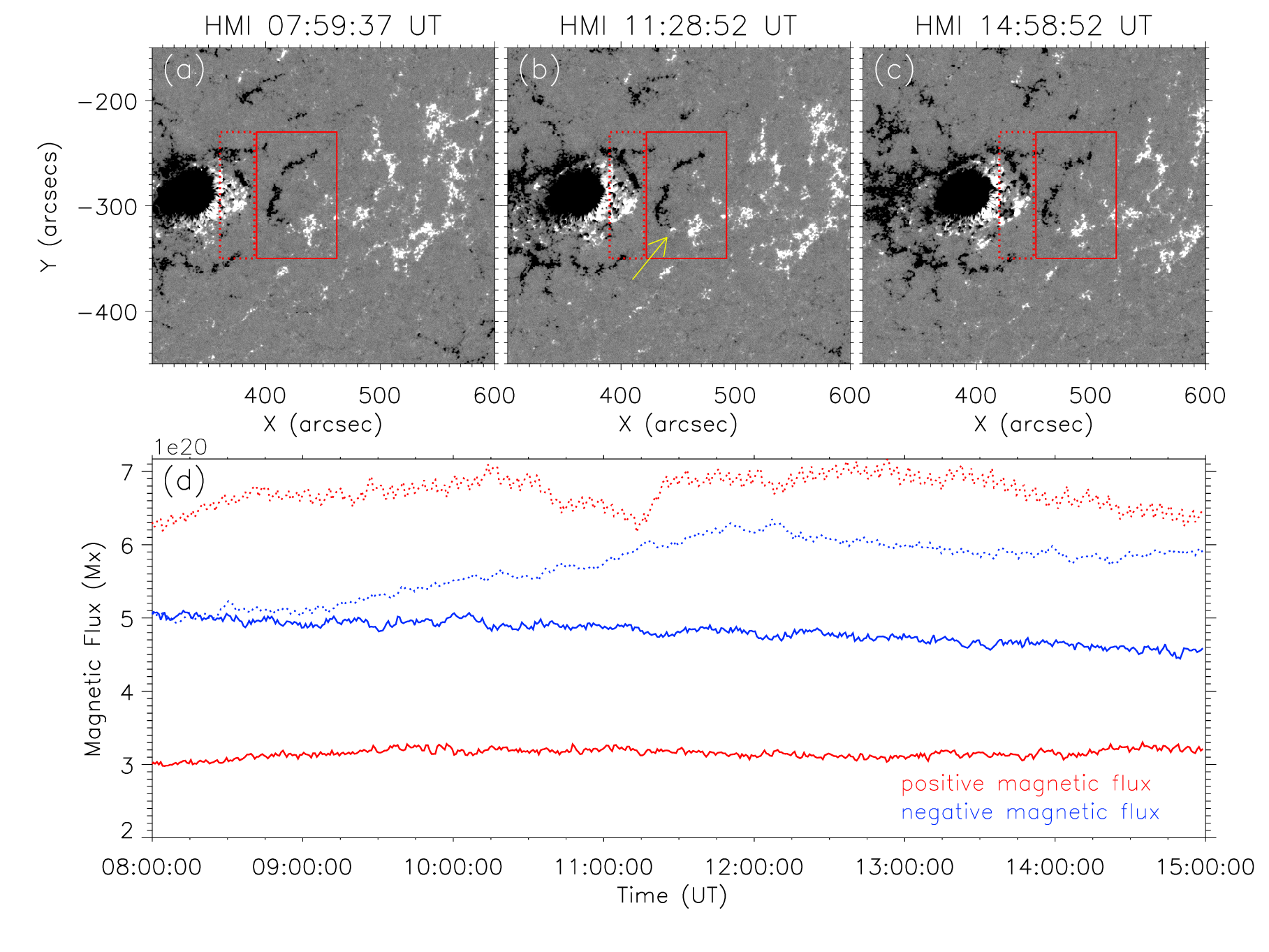}}
\small
        \caption{(a)--(c) HMI magnetograms showing the evolution of the active region during the type I storm. (d) Time profiles of the positive (red) and negative (blue) magnetic fluxes (absolute values). Solid curves represent fluxes integrated within the solid box in the upper panels, and dashed curves correspond to those integrated within the dashed box. An animation accompanies this figure.
                }
\label{Fig5}
\end{figure}

Figure \ref{Fig5} shows the evolution of HMI magnetic field during the type I storm, together with positive and negative magnetic-flux variations measured in selected regions. The active region consists of four main components: a leading negative-polarity sunspot on the left, an adjacent region dominated by positive-polarity moving magnetic features (MMFs: \citealt{1969SoPh....9..347S, 1973SoPh...28...61H, 2005ApJ...635..659H}), a central negative-polarity strip, and an extended, diffuse positive-polarity region on the right.The magnetic flux of the positive MMFs fluctuated slightly (red dashed curve), while that of the negative MMFs first increased slightly and then decreased (blue dashed curve). In the central strip region, the positive magnetic flux remained nearly constant (red solid curve), while the negative magnetic flux decreased continuously (blue solid curve). The accompanying animation reveals continuous emergence and cancellation of small-scale opposite-polarity MMFs, together with persistent contact and cancellation between  central negative-polarity strip and the adjacent right-side positive-polarity region. Figure 5b shows a representative canceling positive-negative magnetic structure. This sustained small-scale photospheric magnetic activity likely drove local magnetic reconnection above the active region.

\begin{figure}%%%%%%%%%%%%%%%%%% FIGURE 6
\centerline{\includegraphics[width=1\textwidth,clip=]{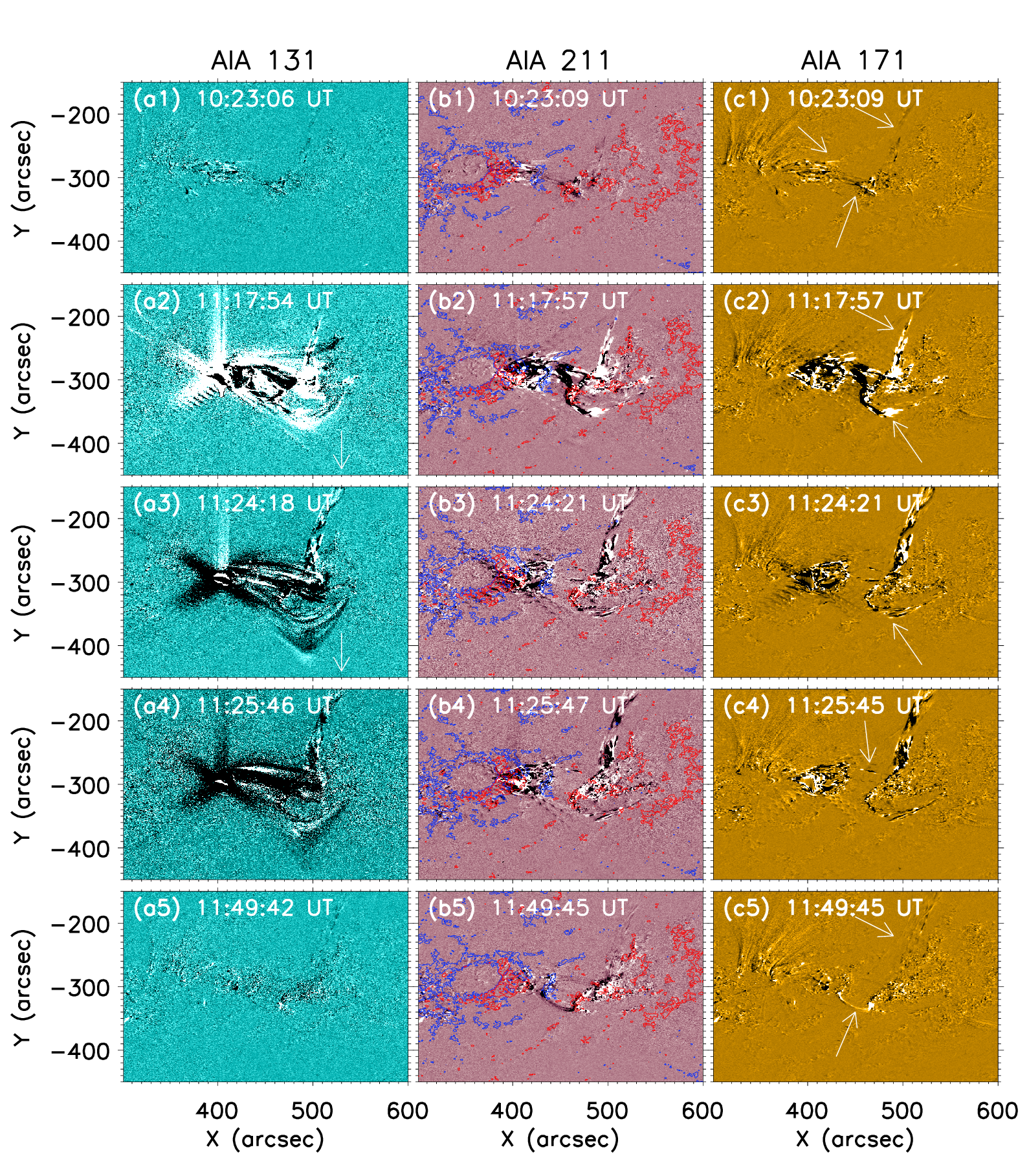}}
\small
        \caption{SDO/AIA 131, 211, and 171 \AA\ running-difference images with a 1 minute time separation, arranged in the left, middle, and right columns, respectively. White arrows mark jets and plasma flows (c1), downward expansion of the loop system possibly associated with loop rise (a2 and a3), enhanced jets and plasma flows along loops (c2 and c3), bidirectional jets (c4), and jets together with a central loop system linking the middle negative-polarity region and the right positive-polarity region (c5). An animation accompanies this figure.
                }
\label{Fig6}
\end{figure}

Figure \ref{Fig6} presents a time sequence of AIA difference images in the 131, 211, and 171 \AA\ passbands. The accompanying animation shows jet activity on both sides of the active region throughout the type I storm, along with persistent plasma flows along the central loop system (panel c1), indicating ongoing magnetic restructuring above the region. Near the onset of the flare and type III bursts, the original loop system expanded downward, possibly in response to the rise of larger-scale magnetic structures (panels a2 and a3). At the same time, jet activity increased, and plasma flows along the right-side loops became more prominent (panels c2 and c3). After about 11:00 UT, magnetic fields on both sides interacted more strongly at higher altitudes and showed signatures of magnetic reconnection, accompanied by stronger jets, including bidirectional jets (panel c4), enhanced flare emission, and type III bursts with bidirectional frequency drifts. During the later phase, after substantial energy release and reduced magnetic stress, the system gradually returned to a more stable configuration. A central loop system connected the middle negative-polarity region to the positive region on the right, while jet activity continued (panel c5). This evolution indicates continued reorganization of magnetic connectivity after the flare.

\subsection{Flare Modulation on the Type I Storm}

\begin{figure}%%%%%%%%%%%%%%%%%% FIGURE 7
\centerline{\includegraphics[width=1\textwidth,clip=]{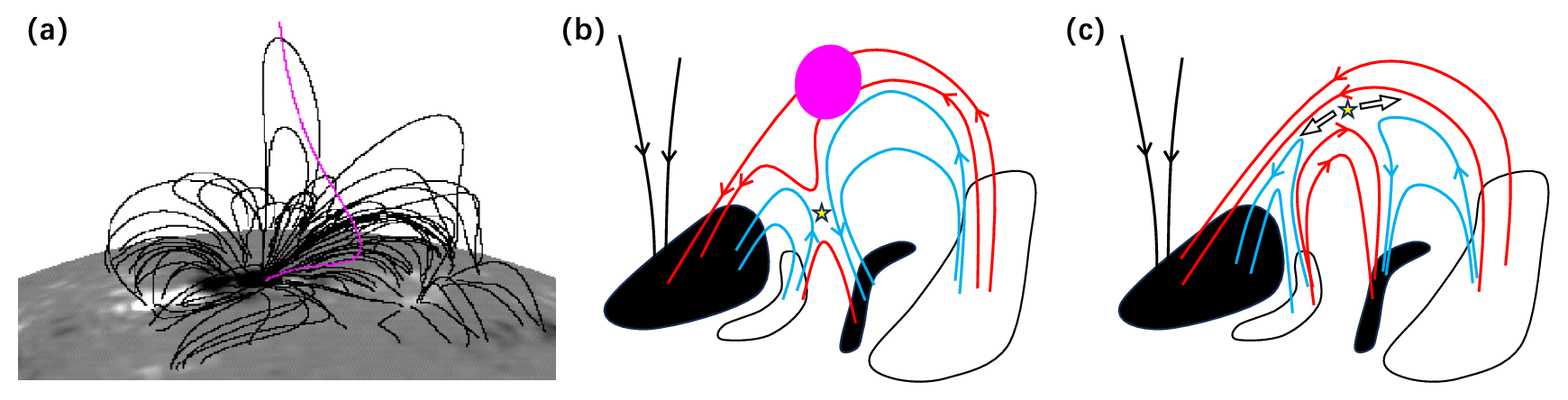}}
\small
        \caption{(a) PFSS extrapolation of the coronal magnetic-field configuration for AR 3529. (b) Schematic illustration of type I storm generation before and after the flare. (c) Schematic of the flare and type III burst generation.
                }
\label{Fig7}
\end{figure}

A potential-field source-surface (PFSS) extrapolation was performed for AR 3529 to infer the large-scale coronal magnetic configuration. Figure \ref{Fig7} presents a schematic scenario for the generation, suppression, and recovery of the type I storm, together with the associated flare and type III bursts with bidirectional frequency drifts.

Before the flare, persistent interaction and cancellation between the MMFs outside the leading sunspot and the central negative strip likely drove small-scale reconnection above the active region. This reconnection accelerated nonthermal electrons and injected them into overlying closed magnetic loops. The trapped electrons sustained the type I storm through either plasma emission or ECME. During this stage, the magnetic topology remained stable, and the type I source position showed little change.

As reconnection proceeded and magnetic stress accumulated, the central arcade rose and compressed the magnetic systems on both sides, and reconnection eventually occurred at higher altitudes (Figure 7c). This process enhanced the flare emission and EUV jets, releasing electron beams both upward and downward from the reconnection region and producing type III bursts with bidirectional frequency drifts. The larger-scale eruptive episode reconfigured the original closed magnetic structures and modified the associated particle-trapping conditions. As a result, the electron confinement that sustained the type I storm was disrupted, leading to a strong suppression of the type I emission.

After the major energy release, the magnetic field above the active region gradually returned to a closed configuration resembling the pre-flare state, allowing the type I emission to resume. However, the height and local density structure of the reformed loops likely differed from those before the flare. The higher-frequency type-I emission after the flare may suggest a source region at lower heights with higher coronal densities.

\section{Summary and Discussion}

Using combined multiwavelength observations, we investigate a type I noise storm on 24 December 2023 and its relation to flares. It was found that (1) The radio source remained stable over an extended interval and showed clear frequency-dependent spatial dispersion. (2) The type I storm was interrupted by an M2.9 flare and later resumed mainly at higher frequencies. (3) Type III bursts with bidirectional frequency drifts and EUV jets appeared near the M2.9 flare, and the starting frequency of the bidirectional drift decreased with time. (4) In the source region beneath, small-scale magnetic elements exhibited persistent emergence, motion, and cancellation, likely driving reconnection at higher altitudes.

Flares and CMEs influence type I storms through multiple mechanisms, as solar radio-burst properties depend on particle acceleration, emission directivity, wave generation, and propagation effects (\citealt{2008JGRA..113.6105L, 2009JGRA..114.2104L}). (1) Type I emission is relatively directive, and large-scale eruptive events such as CMEs can deform coronal magnetic fields, leading to source displacement and radio flux variations (\citealt{1974A&A....32..245C, 2012ApJ...744..167I}). The present event is not accompanied by any CME, and the active region lies near disk center, making emission directivity an unlikely explanation for the observed intensity decrease. (2) Flare-induced changes in temperature, density, and magnetic field can modify wave excitation efficiency and emission intensity (\citealt{1994A&A...291..990Z}), but the type I intensity decrease in this event begins at 10:25 UT, before the C1.9 flare (10:43 UT), the M2.9 flare (11:09 UT), and associated jets, ruling out wave-generation effects. (3) Radio waves cannot propagate when the local plasma frequency exceeds the emission frequency, yet no dense eruptive structures are observed along the propagation path, making propagation effects unlikely. Moreover, the flaring region is nearly co-spatial with the type I source, so the intensity decrease more likely reflects disruption of the confining magnetic structure and changes in particle acceleration.

The observations suggest the following scenario. Persistent small-scale magnetic reconnection supplied energetic electrons, which became trapped in closed magnetic fields and maintained the type I storm. During the M2.9 flare, magnetic reconnection triggered the flare and type III bursts, disrupted the electron confinement required for type I emission, and reconfigured the original closed magnetic system, leading to the suppression of the type I storm. After the eruption, the closed magnetic structure reformed and the type I storm resumed, although its frequency range shifted because of changes in source height or density structure.

The change in source height remains difficult to determine. NRH imaging shows little difference in the projected position of the type I source between the pre- and post-flare phases, although the source position showed oscillatory motion. Such oscillations may arise from ionospheric refraction, which decreases with increasing frequency and usually remains minor above 150 MHz (e.g.,\citealt{1959AuJPh..12..369W, 1981SoPh...73..191D, 1981A&A....96..259B, 2016ApJ...830L...2V, 2021SoPh..296...38L}). As shown in the animation accompanying Figure 2, the systematic frequency-dependent spatial dispersion persisted throughout the event, whereas the source region associated with the type III bursts showed a clear northward displacement during the flare. These signatures more likely reflect intrinsic source evolution than ionospheric refraction.

Bidirectional type III bursts are generally attributed to electron beams released from a magnetic reconnection region and propagating in opposite directions (e.g., \citealt{2000SoPh..194..345R, 2000SoPh..197..375X, 2008Ap&SS.318...87M, 2016ApJ...819...42T}). The decrease in starting frequency may reflect an upward shift of the reconnection region to a lower-density location, a density decrease within the source region itself, or different particle acceleration sites for successive type III bursts. The starting frequencies of 220.3 and 187.8 MHz correspond to electron densities of $\sim$6.0 $\times$ 10$^{8}$ and $\sim$4.4 $\times$ 10$^{8}$ cm$^{-3}$, respectively (assuming fundamental plasma emission), yielding an inferred height difference of $\sim$25000 km from the coronal density models (\citealt{1961ApJ...133..983N, 1999A&A...348..614M}). If the decrease in starting frequency reflects an upward shift of a single reconnection region, the inferred velocity is 900--1000 km s$^{-1}$ over 25 s. This speed is comparable to reconnection outflows or CME-driven shocks. Observations (e.g., of breakout reconnection), however, indicate that the reconnection region moves at only tens of km s$^{-1}$ or remains nearly stationary (\citealt{2012ApJ...760...81K, 2013A&A...555A..40A, 2016ApJ...820L..37C, 2017ApJ...843....8C}), making this interpretation unlikely. Further observations of similar events will help clarify the origin of the starting-frequency variation in bidirectional type III bursts.

%%%%%%%%%%%%%%%%%%%%%%%%%%%%%%%%%%%%%%%%%%%%%%%%%%%%%%%%%%%%%%%%%%%%%%%%%%%
%% Acknowledgements
%
\begin{acks}

We thank the teams of ORFEES, RSTN, GOES, NRH, and SDO for making their data available to us. The authors are grateful to the anonymous referee for the valuable comments.

\end{acks}

\begin{authorcontribution}

Y.L. wrote the main manuscript text, conducted the multiwavelength data analysis, and prepared Figures 1--6. C.L. provided the original idea for the study, performed the PFSS extrapolation and physical interpretation, and contributed to manuscript revision. All authors reviewed the manuscript. Y.T. and N.G. assisted in revising the manuscript and preparing the response to reviewers. Z.L., Z.C. and Y.G. reviewed and edited the final version of the manuscript. Y.W., X.X. and X.H. provided constructive suggestions during the revision stage and approved the final submission.

\end{authorcontribution}

\begin{fundinginformation}

This study is supported by the Scientific Research Fund of Dezhou University (4022504002, 4022504003), the Development Plan for Youth Innovation Teams in Higher Education Institutions of Shandong Province (2025KJG056), the National Natural Science Foundation of China (12103029, 12573033, 12233005), and the Natural Science Foundation of Shandong Province (ZR2021QA079, ZR2024QA169, ZR2025QC1503, ZR2025MS81, ZR2024QA212).

\end{fundinginformation}

\begin{dataavailability}

The ORFEES radio dynamic spectra and Nancay Radioheliograph (NRH) imaging data are publicly available via the Radio Solar DataBase at Nancay (RSDB) at \url{https://rsdb.obs-nancay.fr/}. The RSTN/San Vito radio data are available from the AFRL Products and Data at the NOAA National Centers for Environmental Information (NCEI) at \url{https://www.ncei.noaa.gov/products/space-weather/partners/arfl-products-data}. The SDO/ AIA and SDO/HMI data are available through the Joint Science Operations Center (JSOC) at \url{http://jsoc.stanford.edu/}. The GOES-16 X-ray flux data are from the NOAA NCEI at \url{https://www.ncei.noaa.gov/products}.

\end{dataavailability}

\begin{codeavailability}

The data analysis including PFSS extrapolation in this study was performed using the SolarSoftWare (SSW) IDL package, which is publicly available at https://www.lmsal.com/solarsoft/.

\end{codeavailability}

\begin{ethics}
\begin{conflict}
The authors declare no competing interests.
\end{conflict}
\end{ethics}

\begin{thebibliography}{}


\bibitem[Alissandrakis \emph{et al.}(2015)]{2015A&A...582A..52A}Alissandrakis, C.E., Nindos, A., Patsourakos, S., Kontogeorgos, A., and Tsitsipis, P.: 2015, {\it Astronomy and Astrophysics} {\bf 582}, A52. doi:10.1051/0004-6361/201526265.
\bibitem[Aurass \emph{et al.}(1990)]{1990SoPh..130...19A}Aurass, H., Boehme, A., and Karlicky, M.: 1990, {\it Solar Physics} {\bf 130}, 19. doi:10.1007/BF00156776.
\bibitem[Aurass \emph{et al.}(1993)]{1993SoPh..145..151A}Aurass, H., Hofmann, A., Magun, A., Soru-Escaut, I., and Zlobec, P.: 1993, {\it Solar Physics} {\bf 145}, 151. doi:10.1007/BF00627991.
\bibitem[Aurass \emph{et al.}(2013)]{2013A&A...555A..40A}Aurass, H., Holman, G., Braune, S., Mann, G., and Zlobec, P.: 2013, {\it Astronomy and Astrophysics} {\bf 555}, A40. doi:10.1051/0004-6361/201321111.
\bibitem[Bentley \emph{et al.}(2000)]{2000SoPh..193..227B}Bentley, R.D., Klein, K.-L., van Driel-Gesztelyi, L., D{\'e}moulin, P., Trottet, G., Tassetto, P., and, ...: 2000, {\it Solar Physics} {\bf 193}, 227. doi:10.1023/A:1005218007132.
\bibitem[Benz and Wentzel(1981)]{1981A&A....94..100B}Benz, A.O. and Wentzel, D.G.: 1981, {\it Astronomy and Astrophysics} {\bf 94}, 100.
\bibitem[Bougeret(1981)]{1981A&A....96..259B}Bougeret, J.L.: 1981, {\it Astronomy and Astrophysics} {\bf 96}, 259.
\bibitem[Caroubalos and Steinberg(1974)]{1974A&A....32..245C}Caroubalos, C. and Steinberg, J.L.: 1974, {\it Astronomy and Astrophysics} {\bf 32}, 245.
\bibitem[Chen \emph{et al.}(2016)]{2016ApJ...820L..37C}Chen, Y., Du, G., Zhao, D., Wu, Z., Liu, W., Wang, B., and, ...: 2016, {\it The Astrophysical Journal} {\bf 820}, L37. doi:10.3847/2041-8205/820/2/L37.
\bibitem[Chen \emph{et al.}(2017)]{2017ApJ...843....8C}Chen, Y., Wu, Z., Liu, W., Schwartz, R.A., Zhao, D., Wang, B., and, ...: 2017, {\it The Astrophysical Journal} {\bf 843}, 8. doi:10.3847/1538-4357/aa7462.
\bibitem[Chertok \emph{et al.}(2001)]{2001SoPh..202..337C}Chertok, I.M., Kahler, S., Aurass, H., and Gnezdilov, A.A.: 2001, {\it Solar Physics} {\bf 202}, 337. doi:10.1023/A:1012211412695.
\bibitem[Crosby \emph{et al.}(1996)]{1996SoPh..167..333C}Crosby, N., Vilmer, N., Lund, N., Klein, K.-L., and Sunyaev, R.: 1996, {\it Solar Physics} {\bf 167}, 333. doi:10.1007/BF00146343.
\bibitem[Crosby \emph{et al.}(1997)]{1997ESASP.404..299C}Crosby, N., Vilmer, N., Chiuderi Drago, F., Pick, M., Kerdraon, A., Khan, J., and, ...: 1997, {\it Fifth SOHO Workshop:  The Corona and Solar Wind Near Minimum Activity} {\bf 404}, 299.
\bibitem[Duncan(1981)]{1981SoPh...73..191D}Duncan, R.A.: 1981, {\it Solar Physics} {\bf 73}, 191. doi:10.1007/BF00153154.
\bibitem[Elgar{\o}y(1977)]{elgaroy1977}Elgar{\o}y, {\O}.: 1977, {\it Solar Noise Storms}, Pergamon Press, Oxford. ISBN: 0080210392.
\bibitem[Ginzburg and Zhelezniakov(1958)]{1958SvA.....2..653G}Ginzburg, V.L. and Zhelezniakov, V.V.: 1958, {\it Soviet Astronomy} {\bf 2}, 653.
\bibitem[Guidice \emph{et al.}(1981)]{1981BAAS...13Q.553G}Guidice, D.A., Cliver, E.W., Barron, W.R., and Kahler, S.: 1981, {\it Bulletin of the American Astronomical Society} {\bf 13}, 553.
\bibitem[Hagenaar and Shine(2005)]{2005ApJ...635..659H}Hagenaar, H.J. and Shine, R.A.: 2005, {\it The Astrophysical Journal} {\bf 635}, 659. doi:10.1086/497367.
\bibitem[Hamini \emph{et al.}(2021)]{2021JSWSC..11...57H}Hamini, A., Auxepaules, G., Bir{\'e}e, L., Kenfack, G., Kerdraon, A., Klein, K.-L., and, ...: 2021, {\it Journal of Space Weather and Space Climate} {\bf 11}, 57. doi:10.1051/swsc/2021039.
\bibitem[Harvey and Harvey(1973)]{1973SoPh...28...61H}Harvey, K. and Harvey, J.: 1973, {\it Solar Physics} {\bf 28}, 61. doi:10.1007/BF00152912.
\bibitem[Hey(1946)]{1946Natur.157...47H}Hey, J.S.: 1946, {\it Nature} {\bf 157}, 47. doi:10.1038/157047b0.
\bibitem[Iwai \emph{et al.}(2012)]{2012ApJ...744..167I}Iwai, K., Miyoshi, Y., Masuda, S., Shimojo, M., Shiota, D., Inoue, S., and, ...: 2012, {\it The Astrophysical Journal} {\bf 744}, 167. doi:10.1088/0004-637X/744/2/167.
\bibitem[Kahler \emph{et al.}(1994)]{1994scs..conf..271K}Kahler, S.W., Cliver, E.W., and Chertok, I.M.: 1994, {\it IAU Colloquium 144: Solar Coronal Structures}, 271.
\bibitem[Karpen, Antiochos, and DeVore(2012)]{2012ApJ...760...81K}Karpen, J.T., Antiochos, S.K., and DeVore, C.R.: 2012, {\it The Astrophysical Journal} {\bf 760}, 81. doi:10.1088/0004-637X/760/1/81.
\bibitem[Kerdraon and Delouis(1997)]{1997LNP...483..192K}Kerdraon, A. and Delouis, J.-M.: 1997, {\it Coronal Physics from Radio and Space Observations}, 192. doi:10.1007/BFb0106458.
\bibitem[Lemen \emph{et al.}(2012)]{2012SoPh..275...17L}Lemen, J.R., Title, A.M., Akin, D.J., Boerner, P.F., Chou, C., Drake, J.F., and, ...: 2012, {\it Solar Physics} {\bf 275}, 17. doi:10.1007/s11207-011-9776-8.
\bibitem[Li \emph{et al.}(2008)]{2008JGRA..113.6105L}Li, B., Cairns, I.H., and Robinson, P.A.: 2008, {\it Journal of Geophysical Research (Space Physics)} {\bf 113}, A06105. doi:10.1029/2007JA012958.
\bibitem[Li \emph{et al.}(2009)]{2009JGRA..114.2104L}Li, B., Cairns, I.H., and Robinson, P.A.: 2009, {\it Journal of Geophysical Research (Space Physics)} {\bf 114}, A02104. doi:10.1029/2008JA013687.
\bibitem[Li \emph{et al.}(2017)]{2017SoPh..292...82L}Li, C.Y., Chen, Y., Wang, B., Ruan, G.P., Feng, S.W., Du, G.H., and, ...: 2017, {\it Solar Physics} {\bf 292}, 82. doi:10.1007/s11207-017-1108-1.
\bibitem[Lv \emph{et al.}(2021)]{2021SoPh..296...38L}Lv, M., Chen, Y., Vasanth, V., Radzi, M.S., Abidin, Z.Z., and Monstein, C.: 2021, {\it Solar Physics} {\bf 296}, 38. doi:10.1007/s11207-021-01769-6.
\bibitem[Ma \emph{et al.}(2008)]{2008Ap&SS.318...87M}Ma, Y., Wang, D.Y., Xie, R.X., Wang, M., and Yan, Y.H.: 2008, {\it Astrophysics and Space Science} {\bf 318}, 87. doi:10.1007/s10509-008-9899-z.
\bibitem[Malik and Mercier(1996)]{1996SoPh..165..347M}Malik, R.K. and Mercier, C.: 1996, {\it Solar Physics} {\bf 165}, 347. doi:10.1007/BF00149719.
\bibitem[Mann \emph{et al.}(1999)]{1999A&A...348..614M}Mann, G., Jansen, F., MacDowall, R.J., Kaiser, M.L., and Stone, R.G.: 1999, {\it Astronomy and Astrophysics} {\bf 348}, 614.
\bibitem[Melrose(1970)]{1970AuJPh..23..871M}Melrose, D.B.: 1970, {\it Australian Journal of Physics} {\bf 23}, 871. doi:10.1071/PH700871.
\bibitem[Melrose(1980)]{1980SoPh...67..357M}Melrose, D.B.: 1980, {\it Solar Physics} {\bf 67}, 357. doi:10.1007/BF00149813.
\bibitem[Newkirk(1961)]{1961ApJ...133..983N}Newkirk, G.: 1961, {\it The Astrophysical Journal} {\bf 133}, 983. doi:10.1086/147104.
\bibitem[Ramesh \emph{et al.}(2013)]{2013ApJ...762...89R}Ramesh, R., Sasikumar Raja, K., Kathiravan, C., and Narayanan, A.S.: 2013, {\it The Astrophysical Journal} {\bf 762}, 89. doi:10.1088/0004-637X/762/2/89.
\bibitem[Raulin and Klein(1994)]{1994A&A...281..536R}Raulin, J.P. and Klein, K.-L.: 1994, {\it Astronomy and Astrophysics} {\bf 281}, 536.
\bibitem[Robinson and Benz(2000)]{2000SoPh..194..345R}Robinson, P.A. and Benz, A.O.: 2000, {\it Solar Physics} {\bf 194}, 345. doi:10.1023/A:1005203515701.
\bibitem[Schou \emph{et al.}(2012)]{2012SoPh..275..229S}Schou, J., Scherrer, P.H., Bush, R.I., Wachter, R., Couvidat, S., Rabello-Soares, M.C., and, ...: 2012, {\it Solar Physics} {\bf 275}, 229. doi:10.1007/s11207-011-9842-2.
\bibitem[Schneider(1959)]{1959PhRvL...2..504S}Schneider, J.: 1959, {\it Physical Review Letters} {\bf 2}, 504. doi:10.1103/PhysRevLett.2.504.
\bibitem[Sheeley(1969)]{1969SoPh....9..347S}Sheeley, N.R.: 1969, {\it Solar Physics} {\bf 9}, 347. doi:10.1007/BF02391657.
\bibitem[Spicer \emph{et al.}(1982)]{1982A&A...105..221S}Spicer, D.S., Benz, A.O., and Huba, J.D.: 1982, {\it Astronomy and Astrophysics} {\bf 105}, 221.
\bibitem[Tan \emph{et al.}(2016)]{2016ApJ...819...42T}Tan, B., M{\'e}sz{\'a}rosov{\'a}, H., Karlick{\'y}, M., Huang, G., and Tan, C.: 2016, {\it The Astrophysical Journal} {\bf 819}, 42. doi:10.3847/0004-637X/819/1/42.
\bibitem[Tang \emph{et al.}(2017)]{tang2017ecm}Tang, J.F., Wu, D.J., Zhao, G.Q., and Chen, L. Progress of Electron Cyclotron Maser Emission Mechanism in Solar Physics[J]. Progress in Astronomy, 2017(2): 149--174.
\bibitem[Twiss and Roberts(1958)]{1958AuJPh..11..424T}Twiss, R.Q. and Roberts, J.A.: 1958, {\it Australian Journal of Physics} {\bf 11}, 424. doi:10.1071/PH580424.
\bibitem[Pesnell \emph{et al.}(2012)]{2012SoPh..275....3P}Pesnell, W.D., Thompson, B.J., and Chamberlin, P.C.: 2012, {\it Solar Physics} {\bf 275}, 3. doi:10.1007/s11207-011-9841-3.
\bibitem[Vasanth \emph{et al.}(2016)]{2016ApJ...830L...2V}Vasanth, V., Chen, Y., Feng, S., Ma, S., Du, G., Song, H., and, ...: 2016, {\it The Astrophysical Journal} {\bf 830}, L2. doi:10.3847/2041-8205/830/1/L2.
\bibitem[Wild \emph{et al.}(1959)]{1959AuJPh..12..369W}Wild, J.P., Sheridan, K.V., and Neylan, A.A.: 1959, {\it Australian Journal of Physics} {\bf 12}, 369. doi:10.1071/PH590369.
\bibitem[Wu \emph{et al.}(2012)]{2012PhPl...19h2902W}Wu, C.S., Wang, C.B., Wu, D.J., and Lee, K.H.: 2012, {\it Physics of Plasmas} {\bf 19}, 082902. doi:10.1063/1.4742989.
\bibitem[Wu and Lee(1979)]{1979ApJ...230..621W}Wu, C.S. and Lee, L.C.: 1979, {\it The Astrophysical Journal} {\bf 230}, 621. doi:10.1086/157120.
\bibitem[Xie \emph{et al.}(2000)]{2000SoPh..197..375X}Xie, R.X., Fu, Q.J., Wang, M., and Liu, Y.Y.: 2000, {\it Solar Physics} {\bf 197}, 375. doi:10.1023/A:1026541718129.
\bibitem[Zaitsev \emph{et al.}(1994)]{1994A&A...291..990Z}Zaitsev, V.V., Aurass, H., Kruger, A., and Mann, G.: 1994, {\it Astronomy and Astrophysics} {\bf 291}, 990.




 \end{thebibliography}
\end{document}